\documentclass[%
reprint,
superscriptaddress,
nofootinbib,
amsmath,
amssymb,
aps,
floatfix
]{revtex4-1}

\usepackage{graphicx}
\usepackage{dcolumn}
\usepackage{bm}
\usepackage{hyperref}

\usepackage{textgreek}
\usepackage{mathtools} 
\usepackage{dsfont}
\usepackage{braket}
\usepackage{multirow}
\usepackage{tabularx}
\usepackage[]{algorithm2e}
\usepackage{bigints}

\makeatletter
\renewcommand{\@algocf@capt@plain}{above}
\makeatother

\newcommand{\cosmosis}{\texttt{CosmoSIS}}
\newcommand{\smurves}{\texttt{Smurves}}
\newcommand{\emcee}{\texttt{emcee}}

\newcommand{\pantheon}{\textsc{Pantheon}}
\newcommand{\jla}{\textsc{JLA}}
\newcommand{\des}{\textsc{DES}}

\newcommand{\sn}{SN Ia}
\newcommand{\lcdm}{$\Lambda$CDM}
\newcommand{\wcdm}{$w$CDM}
\newcommand{\kld}{$D_{\textup{KL}}$}

\begin{document}

\raggedbottom



\title{Stress testing the dark energy equation of state imprint on supernova data}

\author{Ben Moews}
\email{bmoews@roe.ac.uk}
\affiliation{Institute for Astronomy, University of Edinburgh, Royal Observatory, Edinburgh EH9 3HJ, UK}

\author{Rafael S. de Souza}
\affiliation{Department of Physics and Astronomy, University of North Carolina at Chapel Hill, NC 27599-3255, USA}

\author{Emille E. O. Ishida}
\affiliation{Universit\'{e} Clermont Auvergne, CNRS/IN2P3, LPC, F-63000 Clermont-Ferrand, France}

\author{Alex I. Malz}
\affiliation{Center for Cosmology and Particle Physics, New York University, 726 Broadway, NY 10004, USA}

\author{Caroline Heneka}
\affiliation{Scuola Normale Superiore, Piazza dei Cavalieri 7, 56126 Pisa, Italy}

\author{Ricardo Vilalta}
\affiliation{Department of Computer Science, University of Houston, 3551 Cullen Blvd., TX 77204-3010, USA}

\author{Joe Zuntz}
\affiliation{Institute for Astronomy, University of Edinburgh, Royal Observatory, Edinburgh EH9 3HJ, UK}

\collaboration{COIN collaboration}
\noaffiliation

\date{\today}

\begin{abstract}
This work determines the degree to which a standard $\Lambda$CDM analysis based on type Ia supernovae can identify deviations from a cosmological constant in the form of a redshift-dependent dark energy equation of state $w(z)$.  
We introduce and apply a novel random curve generator to simulate instances of $w(z)$ from constraint families with increasing distinction from a cosmological constant.
After producing a series of mock catalogs of binned type Ia supernovae corresponding to each $w(z)$ curve, we perform a standard $\Lambda$CDM analysis to estimate the corresponding posterior densities of the absolute magnitude of type Ia supernovae, the present-day matter density, and the equation of state parameter.
Using the Kullback-Leibler divergence between posterior densities as a difference measure, we demonstrate that a standard type Ia supernova cosmology analysis has limited sensitivity to extensive redshift dependencies of the dark energy equation of state.
In addition, we report that larger redshift-dependent departures from a cosmological constant do not necessarily manifest easier-detectable incompatibilities with the $\Lambda$CDM model.
Our results suggest that physics beyond the standard model may simply be hidden in plain sight.
\end{abstract}

\maketitle

\section{Introduction}
\label{sec:intro}

The standard model of cosmology, in which the Universe is composed primarily of cold dark matter (CDM) and a cosmological constant ($\Lambda$), is mainly supported by three observational pillars: 
Big Bang nucleosynthesis (BBN) \cite{Gamow1948}, the cosmic microwave background radiation (CMB) \cite{Dicke1965, Penzias1965, Mather1990, Smoot1992, Spergel2007, Planck2016}, and the discovery of late-time accelerating cosmic expansion \cite{Riess1998, Perlmutter1999, Peebles2003}.

BBN occurred within the first 20 minutes after the Big Bang and is responsible for the production of the lightest nuclides, providing sensitive constraints on the \lcdm\ model (e.g., \cite{Cyburt2016, deSouza2018,deSouza2019}). 
Similarly, the estimated CMB temperature evolution with redshift is corroborated by rotational excitation of molecules and the Sunyaev-Zel'dovich effect \cite{Noterdaeme2011, Luzzi2015}. 
The discovery of accelerated cosmic expansion relies on the observational evidence that type Ia supernovae (\sn) appear fainter than it would be expected in a decelerating universe \cite{Riess1998, Perlmutter1999}. 

The condition for late-time acceleration requires the equation of state parameter of dark energy to be $w < -1/3$, where $w \equiv p / \rho$ is the ratio of its pressure $p$ and energy density $\rho$. 
The postulate of a cosmological constant corresponds to $w = -1$ and has been consistently supported by observational evidence (see, e.g., \cite{Riess2007, Vasey2007, Amanullah2010, Komatsu2011, Sullivan2011, Suzuki2012, Anderson2012} and references therein). 
This constant value is commonly interpreted as a form of vacuum energy in the context of the equation of state of dark energy, the nature of which has garnered the interest of cosmologists for the last two decades \cite{Riess1998, Frieman2008, ORaifeartaigh2018}.
The general notion of a cosmological constant predates the discovery of the accelerating expansion of the Universe (e.g., \cite{Einstein1917, Friedman1922, Lemaitre1927}). 
The concept of dark energy, however, is much broader and has long served as a generic placeholder for the physical cause of an accelerating expansion, which is not necessarily restricted to a constant $w$ (see, e.g., \cite{Frieman2008} for a review).

Typical attempts to probe deviations from the \lcdm\ model assume modifications at the background level, which can be described as a relativistic fluid with an effective time-dependent equation of state. 
The form of the variable equation of state depends on the theory involved, subject to underlying kinetic and potential terms, which can result in considerable variations of $w$ as a function of redshift $z$. 
This also leads to proposals like the Chevallier-Polarski-Linder (CPL) parametrization \cite{Chevallier2001, Linder2003}.
Examples of other non-constant models of dark energy include quintessence \cite{Peebles1988} and, more generally, scalar-tensor theories \cite{Gannouji2006}, with many of them falling under the umbrella of \wcdm\ models \cite{Copeland2006}. 

Theories relying on non-constant parametrizations of $w$ have been tested on real datasets, with no evidence of statistically significant deviations from \lcdm\ being reported \cite{Garnavich1998, Hannestad2004, Chavez2016, Tripathi2017}. 
The same inability to rule out competing theories of dark energy is reported when using \sn\ data under a specialized hypothesis test for ranges of $w$, though future survey data could provide stronger constraints \cite{Genovese2009}.
This competition between a constant and a variable, often redshift-dependent, equation of state is a matter of continuing debate \cite{Huterer2018}. 
A recent example of efforts in testing the CPL parametrization is carried out using the Pan-STARRS\footnote{\url{https://panstarrs.stsci.edu/}} Medium Deep Survey \sn\ data in combination with CMB measurements \cite{Jones2018a, Jones2018b}.

Apart from common parametrizations of $w(z)$ \cite{Jassal2005, deFelice2012}, non-parametric approaches make use of linear or cubic spline interpolation as well as Gaussian processes (GPs) \cite{Zhao2008, Serra2009, Vazquez2012, Hee2017}.
The latter replace the need for placing a limited number of nodes for an interpolation with the choice of a suitable covariance function $K(z, z')$ \cite{Holsclaw2010a, Holsclaw2010b}.
Related research also makes use of non-parametric Bayesian methods based on correlated priors \cite{Crittenden2012}.

Regardless of the preferred representation for the equation of state, the standard analysis consists of including the chosen $w(z)$ model in the supernova likelihood and evaluating the results with the \lcdm\ model as the null hypothesis. 
In this scenario, the goal is to determine which type of behavior is allowed by the data in the context of a given dark energy model, with the prevailing conclusion that currently allowed behaviors are indistinguishable from the \lcdm\ model \cite{Abbott2018d}.

In light of these results, we aim to address the contrapositive question: How robust is a standard \sn\ analysis pipeline to deviations from \lcdm\ in the data? 
We thus investigate whether the traditional \lcdm\ analysis framework is, in this context, a meaningful process to begin with. By creating arbitrary realizations of $w(z)$, we stress-test the viability of currently wide-spread methods to measure $w$ via \sn\ data for the assessment of dark energy models.
To accomplish this goal, we explore current capabilities to discriminate between different models beyond a cosmological constant by running a standard cosmological inference pipeline on random fluctuations of the dark energy parameter $w$ that adhere to physically motivated constraints.

This work is organized as follows: 
The \sn\ mock samples generated for subsequent experiments are described in Section~\ref{sec:data}, along with our procedure for generating data perturbations and the theoretical considerations that have to be taken into account when constraining $w(z)$.
The analysis is performed according to the procedure outlined in Section~\ref{sec:methods}, which provides an overview of the cosmological inference pipeline, the choice of priors, and the measure of posterior differences. 
We present and discuss the results of both the primary investigation and additional experiments for relaxed constraints in Section~\ref{sec:results} and provide our conclusions in Section~\ref{sec:conclusions}.

\section{Data}
\label{sec:data}

\begin{figure}
\includegraphics[width=\columnwidth]{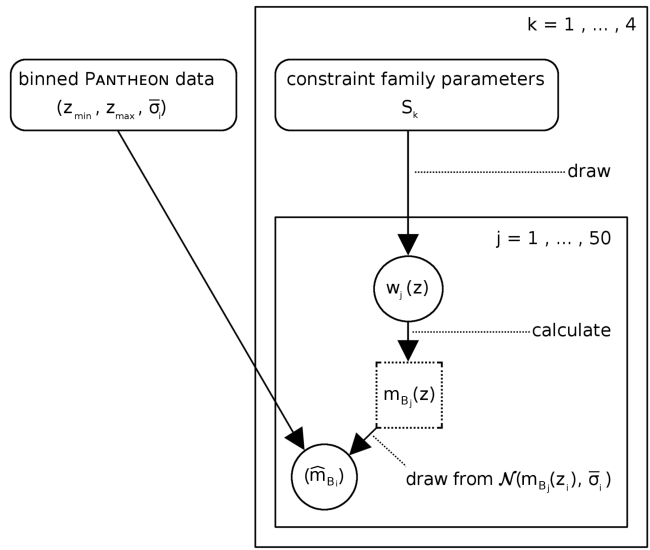}
\caption{
Schematic flowchart of the generation for \pantheon-based \sn\ simulations.
Dotted rectangles denote calculated values, whereas rounded rectangles and circles indicate known values and random variables, respectively. 
Dotted lines mark operations performed at a given point during the process.
}
\label{fig:flow}
\end{figure}

In order to test the limits of a standard \sn\ cosmological pipeline, we generate a series of mock catalogs, each one corresponding to a universe with a different underlying behavior for the dark energy equation of state parameter.
The individual $w(z)$ curves are obtained using a smooth random curve generator described in Section~\ref{sec:smurves}, coupled with physically motivated constraints explained in Section \ref{sec:wz}.
The generated curves are subsequently fed into into a \sn\ simulation pipeline, based on the statistical properties and redshift distribution of the \pantheon\ \sn\ sample \cite{Scolnic2018}. 
Details on our simulation, the process for which is shown in Figure~\ref{fig:flow}, are given in Section~\ref{sec:pantheon}. 

\subsection{Generating perturbations of \texorpdfstring{$\Lambda$}{l}CDM}
\label{sec:smurves}

The construction of mock type \sn\ datasets that can mimic universes with varying dark energy equations of state requires the ability to create $w(z)$ realizations under arbitrarily flexible sets of constraints, for example to define vertical intervals and regulate the maximum number of gradient sign changes.
To this end, we introduce a general-purpose smooth random curve generator that satisfies the need for extensive constraints, together with an easy-to-handle implementation for the wider research community. 
While we use this generator to create realizations of $w(z)$, our method is applicable to a wide array of problems in which generic curves are needed. 
In this context, curve realizations can also be used for function perturbations of arbitrary measurement detail, treating the value at each measurement point as a multiplier for the respective value in a function that is to be smoothly perturbed.

Both node-dependent interpolation approaches and GPs present some significant drawbacks. 
Linear splines lead to sharp changes in the generated functions, while cubic splines are prone to introducing spurious features. 
Similarly, GPs require setting a covariance function and, depending on the kernel, may lack smoothness \cite{Rasmussen2005}. 
In addition, the aforementioned methods hamper the ability to easily subject the generated curves to customized sets of constraints.

To overcome such limitations, we introduce and employ \smurves, a random smooth curve generator that allows for highly customizable and physically motivated constraints to be placed on the curve-generating process. 
The source code of the curve generator, as well as a tutorial and examples, can be found in a public code repository\footnote{\url{https://github.com/moews/smurves}}.
Based on the concept of changes in gravitational direction and magnitude along projectile paths, the generator employs Newtonian projectile motion, adapted to allow for negative values, as the basis for generating curves.  

Given a set of user-specified constraints, \smurves\ generates smooth curves through uniform-random sampling of the number of changes in gravitational direction and the locations of such changes, while adhering to the specified constraints. 
The path is segmented at the sampled change points, and uniform-random samples of the gravitational acceleration are drawn within the bounds of possible curve paths, while respecting the set of interval constraints. 
The method used for curve segment calculations is further summarized, including a pseudocode representation, in Appendix~\ref{app:smurves}.

While we primarily make use of the ability to set intervals and the number of maximum gradient sign changes for this paper, \smurves\ features a variety of additional options that make it applicable to a wider array of problems.
Examples of other capabilities include the use of logarithmic scales and the capacity for perfect convergence in a specified point along the generated curves' paths.

The next section describes the use of \smurves\ to create 50 $w(z)$ curves per constraint family, which imposes boundaries in both dimensions, $z$ and $w$, on each curve sampled at 500 equally-spaced redshift bins on a linear scale. For brevity, we call each such constraint family generated with \smurves\ a ``SmurF''.

\begin{figure}
\begin{center}
\includegraphics[width=\linewidth]{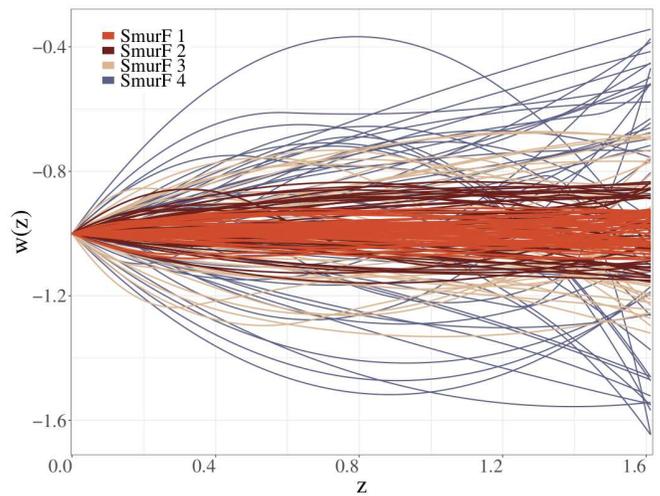}
\caption{
Smooth random $w(z)$ curves generated with \smurves\ to create \sn\ mock observations. 
The figure shows curves from four different constraint families (``SmurFs''), with 50 curves per family, while adhering to a maximum of one gradient sign change for a given curve. 
The varying parameters are the upper and lower boundaries of $w(z)$ for each family.
\label{fig:smurves}
}
\end{center}
\end{figure}

\subsection{Constraints on  \emph{w(z)}}
\label{sec:wz}

We explore families of $w(z)$ curves that evolve within the redshift range covered by the binned \pantheon\ data, $0.0140 < z < 1.6123$, and that are constrained to regions of allowed constant-$w$ models, with a broadest envelope of $-5/3 < w < -1/3$. 

The upper bound of $w = -1 / 3$ is obtained by requiring an accelerated expansion of the Universe at the present time driven by dark energy. 
For each component $i$ of the Universe, this limit corresponds to $\sum_i \left( \rho_i + 3 p_i \right) < 0$, defining the strong energy condition, with equation of state $w_i \equiv p_i / \rho_i$, pressure $p_i$, and energy density $\rho_i$ of energy component $i$ \cite{Dodelson2003}. 
The limit of $w < -1 / 3$ corresponds to a cosmological constant that dominates over other constituents.
The lower bound on $w$ results from the requirement that a so-called Big Rip scenario cannot have occurred within the age of the Universe of roughly one Hubble time $H_0^{-1}$.
The previous term implies that phantom energy, with $w < -1$, becomes infinite in finite time and overcomes all other forms of energy, ripping apart everything, from cosmic structure to atoms, with the Universe ending in a ``Big Rip'' \cite{Caldwell2003}. 
We also note that phantom dark energy violates the null energy condition \cite{Carroll2003}.

While the lowest redshift for the \pantheon\ data is $z = 0.0140$, we set another constraint to let all curves start at $z = 0$ so that $w(0) = -1$. 
This is to agree with near-$z$ cosmological probes bearing small scatter at the lowest redshift bin. 
The resulting set of constrained $w(z)$ curves, shown in Figure~\ref{fig:smurves}, exhibits behaviors that can be found, among others, in effective fluid descriptions of $f(R)$ models \cite{Arjona2018}, scaling, or interacting, dark matter \cite{Chevallier2001}, and bimetric theories of gravity \cite{Koennig2014}.

In practice, this approach means that we evolve the Friedmann equation while including both matter and dark energy as energy components. 
For a flat Universe, this implies
\begin{eqnarray}
H \left( z \right) = H_0 \left[ \Omega_{\mathrm{m}}(1 + z)^{3} + \Omega_{\mathrm{\Lambda}} (1 + z)^{3 \left(1 + w \right)} \right]^{1/2},
\label{eq:hubble}
\end{eqnarray}
where $\Omega_{\mathrm{m}}$ and $ \Omega_{\mathrm{\Lambda}}$ represent the dark matter and dark energy density parameters, respectively. For a flat Universe, we note that $ \Omega_{\mathrm{\Lambda}} = 1 - \Omega_{\mathrm{m}}$.
The current age $t > H_0^{-1}$ of the Universe sets a lower limit on $w$ for a given $\Omega_{\mathrm{m}}$. 
The more negative a phantom component ($w < -1$) is, the faster we reach a Big Rip scenario. 
A lower boundary of $w \gtrsim -2$ corresponds to $\Omega_{\mathrm{m}} = 0.6$, while, for example, $\Omega_{\mathrm{m}} = 0.8$ leads to the requirement $w \gtrsim -2.2$, and $\Omega_{\mathrm{m}} = 0.01$ yields $w \gtrsim -5 / 3$. 
Therefore, we constrain our broadest envelope of $w(z)$ curves to a lower limit of $w = -5 / 3$, conservatively corresponding to a very low matter density and yielding symmetric intervals for the curve limits.

For the three remaining SmurFs, we halve the preceding symmetric interval around $w = - 1$ for each new family, shrinking the allowed envelopes each time to let curves generated from the corresponding families stay closer to the value of the \lcdm\ model. As a result, we generate four curve families with increasing maximum and average deviations from the $\Lambda$CDM model to investigate the degree of compliance for different degrees of compliance with $w(z) = -1$.

We put a final constraint on the curve generator, specifying a maximum number of one for gradient sign changes in the created curves to keep our $w(z)$ curves in line with shapes found in research discussed in Section~\ref{sec:intro}, but explore an increased maximum number of gradient sign changes, as well as the effect of an omission of the $w(z) = 0$ constraint, later in Section~\ref{sec:relaxation}. 

\subsection{\sn\ data simulation}
\label{sec:pantheon}

Observations sensitive to the background expansion such as \sn\ data can be employed to measure the luminosity distance,
\begin{eqnarray}
\label{eq:lumdist}
d_{\mathrm{L}} \left( z \right) = \left( 1 + z \right) d_{\mathrm{H}} \int_{0}^z \frac{\mathrm{d} z'}{E \left( z' \right)},
\end{eqnarray}
where the Hubble distance is $d_{\mathrm{H}} = c / H_0$ and the Hubble parameter is $E(z) = H(z) / H_0$, with $H(z)$ given by Equation~\ref{eq:hubble}.
This is related to the peak B-band magnitude,
\begin{eqnarray}
{m_\mathrm{B}}_i = 5 \log_{10}d_{\mathrm{L}}(z_i) + M,
\end{eqnarray}
of a given supernova $i$ at redshift $z_i$, with absolute magnitude $M$.
We generate \sn\ peak B-band magnitude catalogs by inserting each $w(z)$ curve seen in Figure~\ref{fig:smurves} into Equation~\ref{eq:hubble} and following the process shown in Figure~\ref{fig:flow}.

Our mock data are constructed to mimic the statistical properties and redshift distribution of the \pantheon\ \sn\ sample\footnote{\url{https://archive.stsci.edu/prepds/ps1cosmo/index.html}}, which consists of a total of 1048 \sn\ at redshifts $0.03 < z < 2.3$, representing the largest combined sample of \sn\ observations to date \cite{Scolnic2018}.
We use the publicly available catalog, which is summarized by 40 redshift bins from $z_1 = 0.0140$ to $z_{\mathrm{40}} = 1.6123$.
We note that differences in $w$ between the binned and unbinned versions are smaller than $(1 / 16) \, \sigma$ for statistical measurements \cite{Scolnic2018}, which makes this an adequate and easy-to-handle data representation for a large number of analysis pipeline runs.

\begin{figure}
\includegraphics[width=\columnwidth]{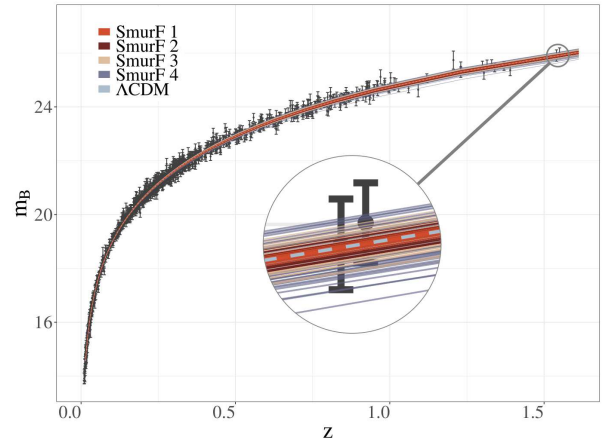}
\caption{
Peak B-band magnitudes $m_\mathrm{B}$ as a function of redshift $z$ for different dark energy equation of state ($w(z)$) realizations. 
The figure shows the diagrams for the \lcdm\ model (dashed line), as well as 50 random $w(z)$ curves for each of the four constraint families, which represent increasing deviations from \lcdm. 
Black points depict the \pantheon\ dataset and respective uncertainties, and the insets highlight $w(z)$ models regarding \lcdm\ as mostly falling within the data uncertainty, even at redshifts as high as $z \gtrsim 1.5$.
}
\label{fig:hubble}
\end{figure}

We propagate the curves through a simulation pipeline using \cosmosis, as described in Section~\ref{sec:cosmosis}. 
The simulation pipeline also takes into account the full covariance matrix, which includes effects due to photometric error, the uncertainty in the mass step correction, uncertainty from peculiar velocity and redshift measurement, distance bias correction, and uncertainty from stochastic lensing and intrinsic scatter.
Peak B-band magnitudes for $w(z)$ curves are shown in Figure~\ref{fig:hubble} to demonstrate the similarity of results even at high redshifts.

\section{Methods}
\label{sec:methods}

We run a full analysis pipeline that assumes a constant-$w$ dark energy model, hereafter called $\Psi_{w_{\mathrm{const}}}$, to infer the posterior probability distribution of $w$, $\Omega_{\mathrm{m}}$, and $M$ as described in Section~\ref{sec:cosmosis}. 
In Section~\ref{sec:priors}, we list and justify our choice of priors for parameters.
Finally, in Section~\ref{sec:metric}, we introduce the metric by which we compare simulation-based posteriors and those from real \sn\ \pantheon\ data.

\subsection{Pipeline with \cosmosis}
\label{sec:cosmosis}

\cosmosis\ is a cosmological parameter estimation code \cite{Zuntz2015}, which models cosmological likelihoods and calculations as a sequence of independent modules that read and write their inputs and outputs to a central data storage block. 
The package has been used extensively for parameter estimation by the Dark Energy Survey (\des) (e.g., \cite{Abbott2018a, Abbott2018b, Abbott2018c, Elvin-Poole2018, Troxel2018}), among others \cite{Barreira2015, Harrison2016, Krause2017, Lin2017}.

We utilize two \cosmosis\ pipelines; the first simulates data using the $w(z)$ realizations described above, and the second analyzes the simulated data using the \emcee\ sampler \cite{Goodman2010, Foreman-Mackey2013} under a standard cosmological model. 
The process of \emcee\ is described in detail in Appendix~\ref{app:emcee}.

We connect these two pipelines in a script to iterate the process over the curves from each SmurF using four standard library modules: \texttt{consistency}, which computes the complete set of cosmological parameters, \texttt{camb} \cite{Lewis1999}, which, in our case, calculates cosmological background functions, and \texttt{pantheon}, which computes the \pantheon\ likelihood. 
A custom module is used to read in tabulated $w(z)$ functions and cast them to the form used in \texttt{camb}.

\begin{figure*}
\begin{center}
\includegraphics[width=0.95\linewidth]{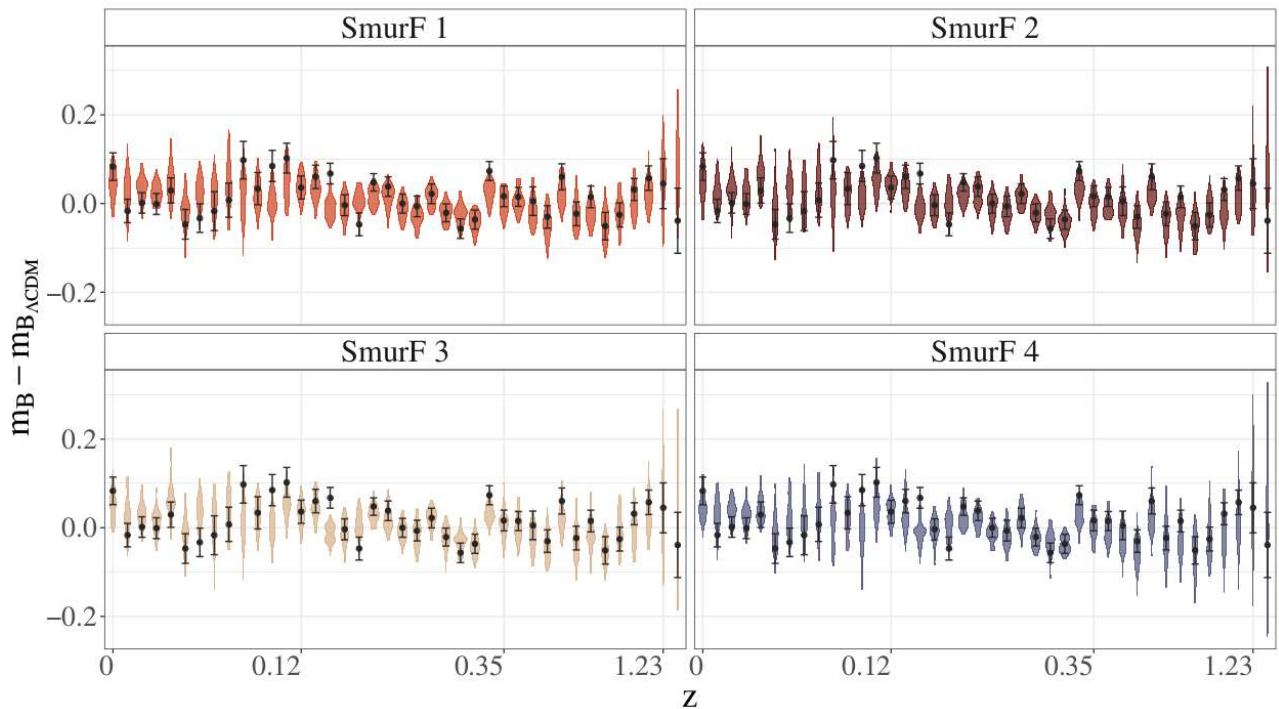}
\caption{
Visualization of peak B-band magnitude ($m_\mathrm{B}$) residuals between our simulated data and \lcdm, as well as between observed \pantheon\ data and the \lcdm\ model. 
In both cases, \lcdm\ corresponds to $\Omega_{\mathrm{m}} = 0.307$ and $M = -19.255$. 
The violin plots for each of the 40 redshift ($z$) bins show a rotated kernel density plot of the distributions of values for each of 50 different realizations for one SmurF per panel.
Black dots indicate binned \pantheon\ data, with vertical black lines representing the error bars of one standard deviation.
The comparison is plotted as the difference between the respective peak B-band magnitudes and expected $\Lambda$CDM values, $m_\mathrm{B} - m_\mathrm{B}{_{\Lambda \mathrm{CDM}}}$, to show both the deviation from theoretical values and the distributions of simulated \sn\ data around observed values.
\label{fig:distmods}
}
\end{center}
\end{figure*}

For Gaussian likelihoods, \cosmosis\ automatically generates simulated outputs incorporating both the signal based on the used model and noise, as described in Appendix~\ref{app:cosmosis}. 
Employing the reported uncertainties on $m_\mathrm{B}$ and the full covariance matrix, we use this process to simulate peak B-band magnitudes at the same redshift values as reported for the real data in the binned \pantheon\ sample. 
The distributions of these mock peak B-band magnitudes are provided in Figure~\ref{fig:distmods}.

\subsection{Choice of priors}
\label{sec:priors}

\begin{table}
\centering
\caption{Priors for the estimation of cosmological and nuisance parameters. 
U$(\cdot)$ denotes a uniform distribution, whereas we use ``fixed'' to indicate a Dirac delta function with $\delta(x) = \infty$ for an $x$ from the column of initial values.}
\label{tab:priors}
\begin{tabularx}{\columnwidth}{l*{2}{>{\centering\arraybackslash}X}}
\hline
\hline
Parameter & Prior & Initial value \\
\hline
$\Omega_\mathrm{m}$ & U(0.01, 0.6) & 0.307\\ 
$M$ & U(-20.0, -18.0) & -19.255\\ 
$w$ & U(-2.0, -0.3333) & -1.026\\
$\Omega_\mathrm{k}$ & fixed & 0\\
$\Omega_\mathrm{b}$ & fixed & 0.04\\ 
$h_0$ & fixed & 0.7324 \\
\hline
\hline
\end{tabularx}
\end{table}

We vary our cosmology via the present-day matter density $\Omega_{\mathrm{m}}$ and the dark energy equation of state $w$.
We assume a flat Universe with $\Omega_{\mathrm{k}} = 0$ and, therefore, a dark energy density of $\Omega_{\mathrm{\Lambda}} = 1 - \Omega_{\mathrm{m}}$. 
We keep the present-day Hubble parameter fixed to $h_0 = 0.7324$ \cite{Riess2016}, and the cosmic baryon density to $\Omega_{\mathrm{b}} = 0.04$ \cite{Cooke2013}. 
An additional nuisance parameter is the absolute magnitude of \sn\ $M$, which is degenerate with the Hubble parameter. 

Our set of estimated parameters from the \emcee\ sampler is $\{ \Omega_{\mathrm{m}}, w, M \}$. 
We choose uniform priors for all parameters, with bounds given in Table~\ref{tab:priors}. 
The range for the absolute magnitude $M$ encompasses previous constraints given, for example, by the SDSS-II/SNLS3 Joint Light-Curve Analysis (\jla) \cite{Betoule2014}. 
The central starting value of $M = -19.255$ is chosen from a preliminary maximum likelihood run with \pantheon\ data. 
The prior over $\Omega_{\mathrm{m}}$ covers allowed parameter ranges as estimated by present-day \sn\ samples like \jla\ and \pantheon. 
The starting point for the dark matter parameter is $\Omega_{\mathrm{m}} = 0.307$, which corresponds to the \pantheon\ \wcdm\ best-fit value. 
Analogously, the central value for $w$ is set to $w = -1.026$ \cite{Scolnic2018}. 

The prior range on $w$ coincides with the allowed values for the families of $w(z)$ curves considering the prior upper bound of $\Omega_{\mathrm{m}} = 0.6$ (see Section~\ref{sec:wz} for a detailed description of the allowed $w$-interval). 
For our parameter estimation, we loosen the symmetric lower-bound requirement, with $w = -2$ as our lower limit to cover the allowed upper boundary of $\Omega_{\mathrm{m}}$ from \sn\ at $3 \, \sigma$. 

\subsection{Comparison criteria}
\label{sec:metric}

Conventional error contours, used ubiquitously in cosmology, are estimated from samples from posterior probability distributions $p(\theta | D, \Psi)$ of parameters of interest, in our case $\theta = \{ \Omega_{\mathrm{m}}, w, M \}$, conditioned on the cosmological model $\Psi$ and data $D = \{d_{i}\}_{N}$, where $i$ runs over the number $N$ of observations. 
For \pantheon, the data is presented as $D_{\pantheon} = \{z_i, {m_\mathrm{B}}_i, \sigma_{m_\mathrm{B},i} \}_{40}$ for bins $i$.

We consider each individual $w(z)$ curve separately, but group them by constraint family $S_k$, as depicted in Figure~\ref{fig:flow}, for interpretability (see Appendix~\ref{app:math} for a detailed justification of this procedure).
For $j \in \{ 1, 2, \dots, 50 \}$, each of 50 simulated data sets $D_j$ is generated with the curve $w_j(z)$, and our experimental design yields samples from the posteriors $p_j \equiv p(\theta | D_{j}, \Psi_{w_{\mathrm{const}}})$.
Each posterior corresponds to the probability of parameters from a cosmological model $\Psi_{w_{\mathrm{const}}}$ conditioned on the data generated from $w_j(z)$. 
We also apply the same pipeline to 50 realizations of the data under the \lcdm\ model, producing $p_{\Lambda_j} \equiv p(\theta | D_{{\Lambda\mathrm{CDM}}_j}, \Psi_{w_{\mathrm{const}}})$, and to the real \pantheon\ data, producing $p_{\pantheon} \equiv p(\theta | D_{\pantheon}, \Psi_{w_{\mathrm{const}}})$. 

To compare the samples from each mock universe to their \lcdm\ counterparts, we adopt a measure suited to quantifying the difference between probability distributions. 
The Kullback–Leibler divergence (\kld) \cite{Kullback1951},
\begin{eqnarray}
\label{eq:kld}
D_{\mathrm{KL}} = \int_{-\infty}^{\infty} p(x) \ln\left[\frac{p(x)}{\hat{p}(x)}\right] \mathrm{d}x,
\end{eqnarray}
is the directional difference between a reference probability distribution $p(x)$ and a proposed approximating probability distribution $\hat{p}(x)$.
The \kld\ has been applied within astronomy only to a limited extent, but is gaining popularity \cite{Kilbinger2010, Ben-David2015, deSouza2017, Hee2017, Malz2018, Nicola2018}.

Unlike symmetric measures of the distance between two probability distributions, such as the familiar root-mean-square-error, the \kld\ is defined as the directional loss of information due to using an approximation in place of the truth; 
we must designate one distribution as a reference from which the proposal distribution diverges. 
A generic example of a pair of reference and proposal distributions can be defined by posterior samples derived from a large set of observations, as opposed to posterior samples derived from a small subset thereof. 
There is, therefore, an implicit assumption that the former is closer to the truth than the latter, which may be an approximation when the rest of the observations are unavailable.

In our case, the samples from $p_{\pantheon}$ always serve as the reference distribution, and the samples from $p_j$ and $p_{\Lambda_j}$ always act as the proposal distribution.

\section{Results and Discussion}
\label{sec:results}

In the previous sections, we describe both the data and our methodology. 
In Section~\ref{sec:primary}, we present the results of primary experiments, together with a discussion of the underlying causes and implications for \sn\ investigations. 
In addition, we relax the different constraints for two of the constraints families in Section~\ref{sec:relaxation} to explore the impact such changes have on the resulting \kld\ distributions. 
In the first of these two additional experiments, we generate $w(z)$ curves with an increased maximum number of gradient sign shifts, whereas the second experiment eliminates the requirement that $w(z) = -1$.

\subsection{Primary experiments}
\label{sec:primary}

For each SmurF, as described in Section~\ref{sec:smurves}, we generate 50 $w(z)$ curves that are fed into the \cosmosis\ simulation and analysis pipeline described in Section~\ref{sec:cosmosis}. 
This results in four sets of 50 posterior distributions for parameters $\{ \Omega_{\mathrm{m}}, w, M \}$, or $p_{S_{k},j}$, where $k \in \{1,2,3,4\}$ identifies the SmurF and $j \in \{1, 2, \dots, 50\}$ denotes its realizations (see Section \ref{sec:metric} for details on notation). 
In addition, 50 datasets from a \lcdm\ model are generated to illustrate the impact allowed by current statistics and systematic uncertainties. 
We feed these simulations, as well as the original binned \pantheon\ dataset, into the same analysis pipeline. 
Posteriors derived from all simulated data are then compared to the \pantheon\ results using the Kullback-Leibler divergence \kld, described in Section \ref{sec:metric}.
 
\begin{figure}
\begin{center}
\includegraphics[width=\linewidth]{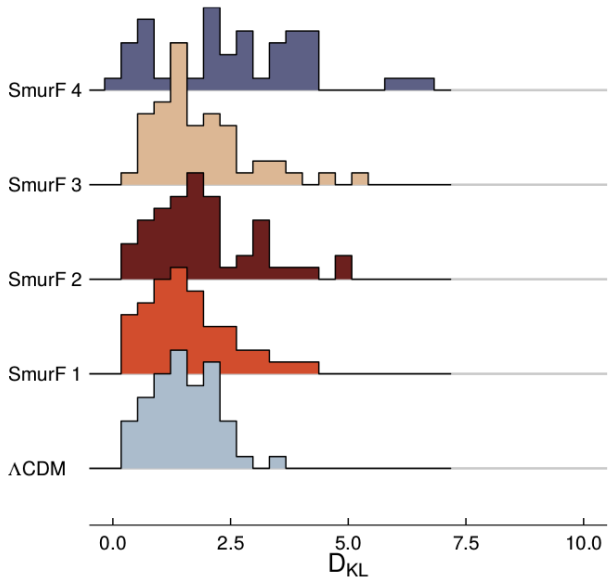}
\caption{
Histograms of the Kullback-Leibler divergence (\kld) for different sets of constraints. 
The shown histograms depict the distribution of \kld\ values for the \lcdm\ case and each SmurF used to generate simulated \sn\ peak B-band magnitudes.
\kld\ values are calculated for the posterior distributions of parameters obtained through a standard \lcdm\ analysis pipeline that considers only constant $w$ models. 
\label{fig:klds}
}
\end{center}
\end{figure}
 
Figure~\ref{fig:klds} shows histograms of \kld\ values for each SmurF along with those from \lcdm\ simulations. 
In accordance with our expectations, the distributions of \kld\ values for constraint families with increasingly wider $w$-intervals, from SmurF 1 through 4, show a systematic shift towards higher means, larger variances, and multimodality. 
These differences are, however, small enough that the bulk of \kld\ values for each SmurF coincides with the \kld\ range covered by the \lcdm\ case, presenting a serious obstacle for the detection of deviations from a cosmological constant.

\begin{figure*}[t]
\begin{center}
\includegraphics[width=0.95\linewidth]{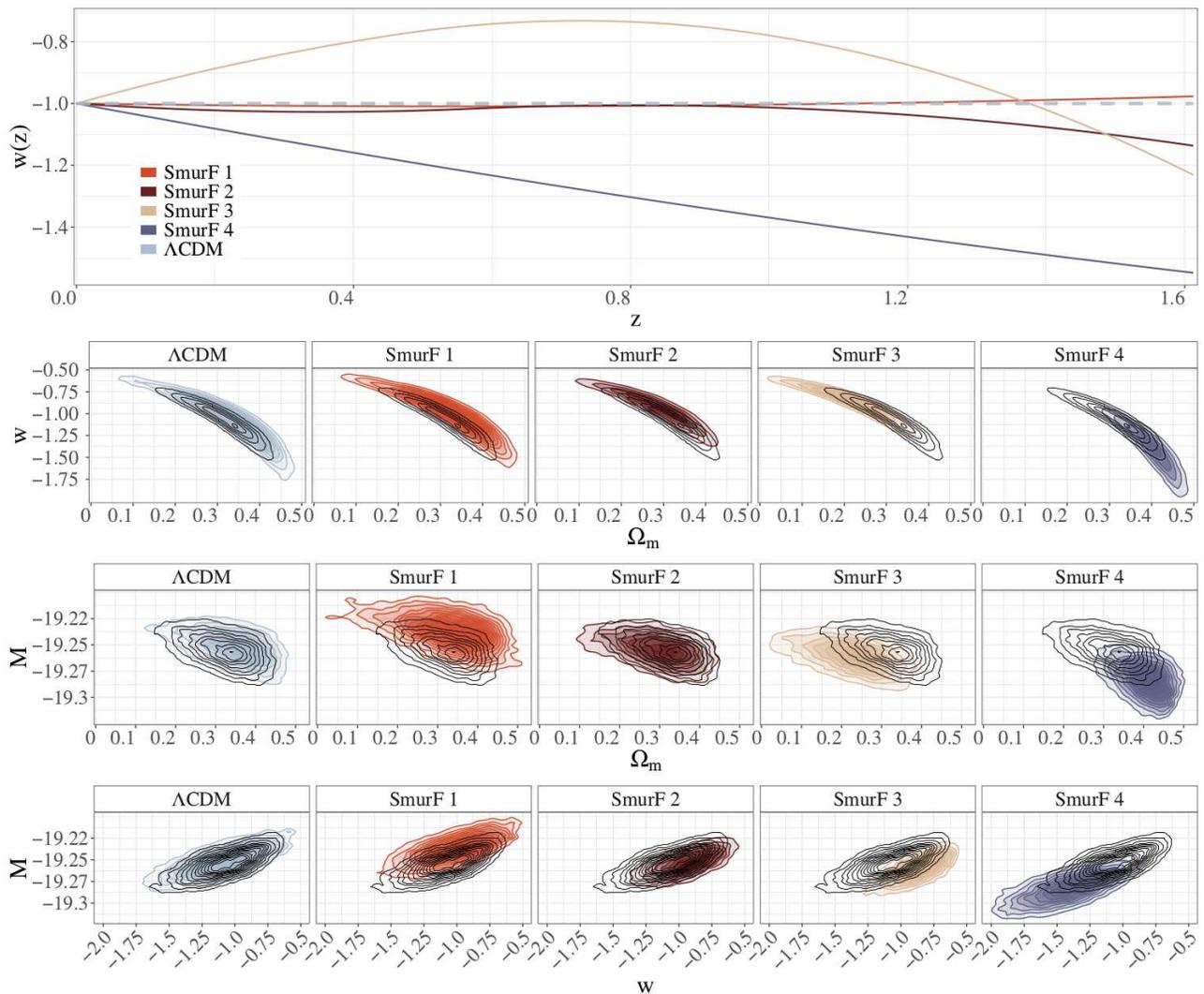}
\caption{
First row: Representative redshift-dependent dark energy equation of state ($w(z)$) curves associated with the median \kld\ per constraint family (full lines) and the \lcdm\ case (dashed line). 
Second row: Posteriors for $w$ and dark matter density $\Omega_{\mathrm{m}}$ per constraint family. 
The four plots depict the posterior distributions for the above-mentioned curves (colored contours), as well as the posteriors for the \pantheon\ analysis case (black contours). 
Third and fourth row: With $M$ as the absolute magnitude, the plots show two-dimensional posteriors for $M \times \Omega_{\mathrm{m}}$ and $M \times w$, respectively. 
}
\label{fig:medianexample}
\end{center}
\end{figure*}

This effect is better visualized by a representative $w(z)$ function for each SmurF and the respective posteriors, shown in each column of Figure~\ref{fig:medianexample}. 
The top row shows $w(z)$ curve associated with the median \kld\ value for each SmurF, as well as the constant $w = -1$ line. In doing so, we enable the comparison of single representative curves, which we also visualize to ensure that deviations from $w(z) = -1$ in representatives follow the same progression toward larger deviations as the increasing deviations distinguishing different SmurFs.

Each curve approximately covers the allowed $w$ intervals of its respective constraint family, thus confirming the applicability of a median-\kld\ approach for choosing a representative SmurF instance. 
The bottom three rows show two-dimensional posterior distributions, for parameters $\{\Omega_{\mathrm{m}}, w, M\}$, for each SmurF and the \lcdm\ case (colored contours) superimposed on the posteriors from \pantheon\ data (black contours). 
Similarly, posterior distributions from the \lcdm\ model, together with SmurFs 1, 3, and 4, go from agreement to disagreement with \pantheon. 
Posteriors from SmurF 2, on the other hand, show an unexpected visual match with both real \pantheon\ data results and the \lcdm\ case, despite its associated $w(z)$ exhibiting larger deviations from $w = -1$ than the one associated with SmurF 1.
Notably, the representative curve from SmurF 2 features larger deviations from the $\Lambda$CDM case than the representative curve from SmurF 1 in both low-$z$ and high-$z$ regimens, meaning that larger deviations from the $\lambda$CDM case do not necessarily result in posteriors considerably different from the ones produces by $w(z) = -1$.

This apparent discrepancy between notable inconsistencies in $w(z)$ and compliant posterior estimates derives from the fact that, while $w(z)$ can change widely, the observable signature of $w(z)$ relies on the peak B-band magnitude $m_\mathrm{B}$.
The dependence of $m_\mathrm{B}$ on the integral of the Hubble parameter leads to a statistical degeneracy that makes such posteriors indistinguishable from \lcdm\ within the current magnitude precision level and probed redshift range. 
Coupled with the large \kld\ overlap between SmurF instances and \lcdm\ results seen in Figure~\ref{fig:klds}, this directly extends to a considerable chance of mistaking an equation of state varying significantly with redshift for one in reasonable agreement with a cosmological constant.

\begin{figure}
\begin{center}
\includegraphics[width=\columnwidth]{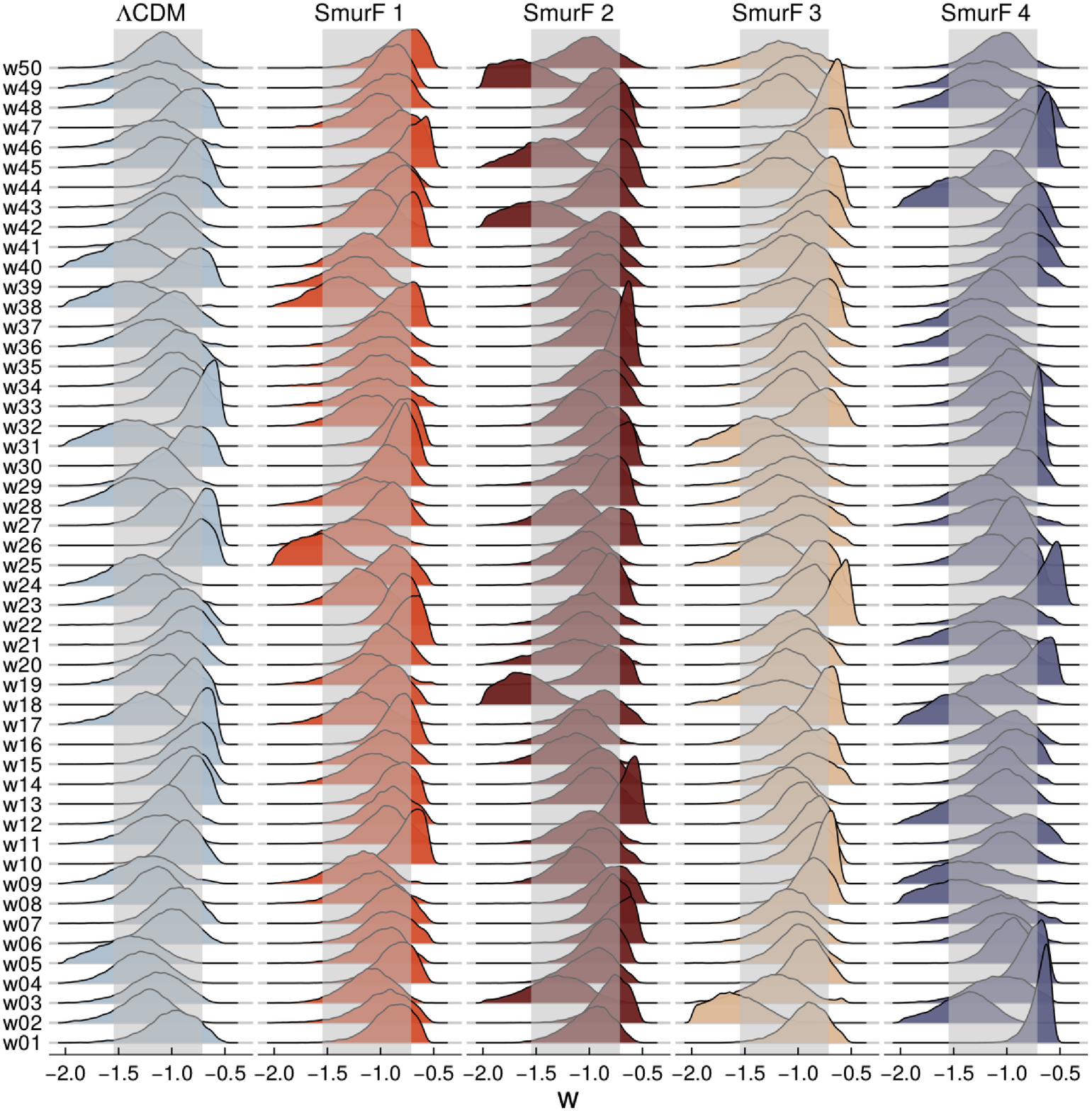}
\caption{
Ridgeline plots for the dark energy equation of state parameter $w$. 
Each row depicts the posterior densities of $w$ for all 50 curves, for each of the four constraint families as well as the simulations for the \lcdm\ case. 
The transparent bands covering the middle section of each column show the 95\% credible interval for the \pantheon\ sample, analyzed under a constant-$w$ model. 
}
\label{fig:familydensities}
\end{center}
\end{figure}

A more detailed view of all posteriors over $w$ is shown in the ridgeline plots of Figure~\ref{fig:familydensities}, in which the means, as well as the bulk of the probability, fall within the 95\% credible intervals of the \pantheon\ results under a constant-$w$ hypothesis. 
SmurF 2, in particular, shows more constrained posteriors, which offers an explanation for the agreement of the median-\kld\ representative's posterior with the \lcdm\ case. 
It does, however, also feature four obvious outliers reaching far beyond the left boundary of the credible interval, which demonstrates the variability in the agreement of $w$-posteriors within the same constraint family.

Naturally, all of the the aforementioned results are bounded by the \pantheon-like quality of our simulations. 
Current surveys such as \des\ continue to contribute to the number of \sn\ observations \cite{Abbott2018a}. 
Though the \des\ \sn\ samples used in combination with additional external samples amount to less than a third of \pantheon's sample size, \des\ results indicate smaller intrinsic scatter in the Hubble diagram, taking one step further in the attainment of higher-quality \sn\ samples \cite{Brout2018}.
These new and future datasets will certainly increase our ability to discriminate between different models for the dark energy equation of state parameter. 

It is, however, important to highlight the non-intuitive and unavoidable behavior derived from the nature of distance measurements as an integral over the Hubble parameter. 
Given a dataset with sufficiently low measurement and systematic uncertainties, especially at high redshifts, discrimination between phenomenologically close models is possible, but we cannot rely on the assumption that substantial redshift-dependent changes in $w(z)$ will necessarily result in detectable biases under a constant-$w$ analysis. 
This is especially the case for \sn-only analyses~\cite{Miranda2017, Zhao2017, Abbott2018d, LHuillier2018}.

Caution should be exercised in using other cosmological observables to break the degeneracy via constraining additional parameters.
This strategy is wide-spread in the literature, to the point that recent research questions the use of \sn\ data without such additional observables \cite{Peracaula2018}.
It is, however, important to keep in mind that supernovae are the primary dynamical observable that probe the line of sight directly, and consequently impose boundaries in the behavior of $w$. 
The use of additional probes such as weak lensing can, with insufficient information on the baryonic physics involved, introduce new biases, for example in the CPL parametrization \cite{Copeland2018}.

In summary, we recognize the need to combine complementary observables, for example baryon acoustic oscillations and CMB data, while making use of careful statistical analyses capable of probing more subtle behaviors of the dynamical evolution of dark energy. 
Although paramount for a more general discussion of this topic, the addition of extra observables exceeds the scope of this paper.

\subsection{Relaxed constraints on \emph{w(z)}}
\label{sec:relaxation}

\begin{figure}
\begin{center}
\includegraphics[width=\linewidth]{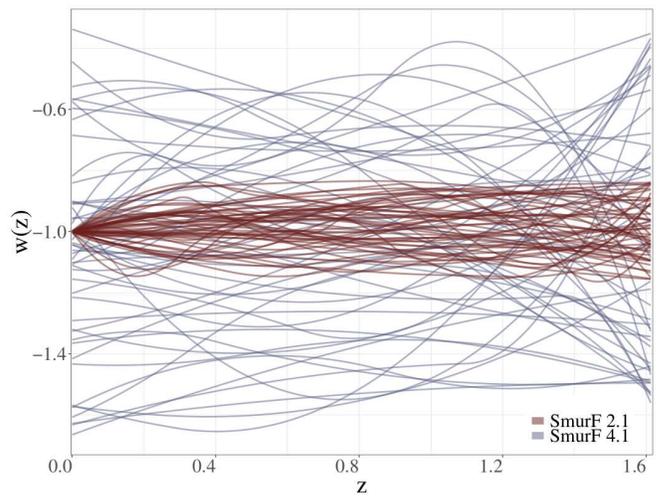}
\caption{
Smooth random dark energy equation of state ($w(z)$) curves generated with \smurves\ to create mock \sn\ observations for additional experiments.
The figure shows curves from two different constraint families, SmurF 2.1 and SmurF 4.1, with 50 curve realizations per family.
\label{fig:curves2}
}
\end{center}
\end{figure}

In a bid to push our analysis a bit further, we relax the constraints put on the curve generator for SmurFs 2 and 4 for illustrative purposes. 
For SmurF 2, we increase the maximum number of gradient sign changes from one to 10, allowing for more complicated functions to be realized. 
In contrast, for SmurF 4, we omit the requirement that $w(0) = - 1$ to allow curves to start at arbitrary values within the allowed $w(z)$ interval. 
The respective curves used in these additional experiments are depicted in Figure~\ref{fig:curves2}.

\begin{figure}
\begin{center}
\includegraphics[width=\linewidth]{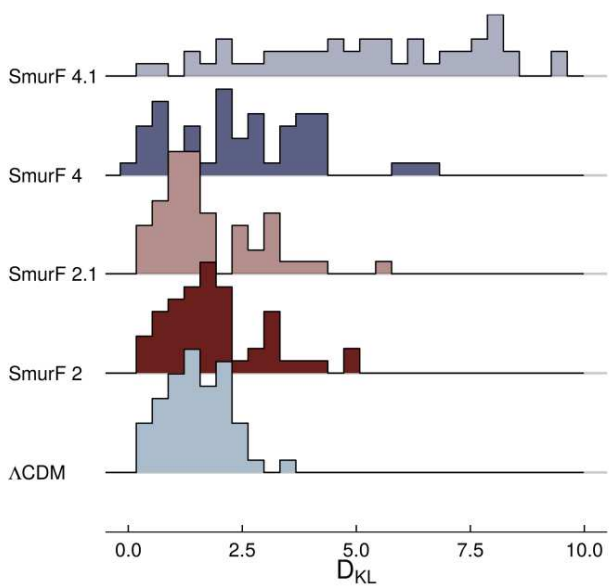}
\caption{
Histograms of the Kullback-Leibler divergence (\kld) for different constraint families. 
The histograms show the distributions of \kld\ values, with a total of 50 redshift-dependent dark energy of state curves $w(z)$ per family. 
In doing so, this figure facilitates the comparison of two previous constraint families, SmurF 2 and SmurF 4, with further relaxed constraint families, namely SmurF 2.1 and SmurF 4.1, as well as with the \lcdm\ case.
\label{fig:klds_extra}
}
\end{center}
\end{figure}

To assess the impact of these further constraint relaxations, their \kld\ distributions are shown in Figure~\ref{fig:klds_extra}, along with those from SmurF 2, SmurF 4, and the \lcdm\ case.
The \kld\ distribution of SmurF 2.1 still holds the same overall shape of SmurF 2 and occupies a range of \kld\ values between those covered by SmurF 2 and 4. 
This demonstrates that the use of more complicated functions, for example the larger maximum number of gradient sign changes in SmurF 2.1, has a lesser impact than simpler functions allowed to vary in a larger interval, as is the case for SmurF 4.1, when constrained to the same $w(z)$ intervals and initial conditions.
The complexity of $w(z)$ curves does, as a result, seem to have less of an effect on distinguishability than the intervals in which they live. This is, again, a consequence of the dependence of $m_\mathrm{B}$ on the integral over the Hubble parameter, meaning that faster variations in $w(z)$ tend to be smoothed out observationally. 
Residual additional variations, which are still present, lead to the slightly higher spread in the corresponding \kld\ distribution.

When we omit the $w(0) = -1$ constraint, which restricts generated $w(z)$ curves to exhibit stark variations from the \lcdm\ case at very low redshifts, we find ourselves confronted with a very different result. 
Relative to SmurF 4, SmurF 4.1 exhibits larger \kld\ values with a considerably wider spread.
We also note that the distribution of \kld\ values is much flatter than for distributions constrained to $w(0) = -1$, without a peak at low \kld\ values. 
This wider spread and flattened distribution can be attributed to introducing an offset in our observable $m_\mathrm{B}$, since $m_\mathrm{B}$ averages over $w(z)$ via the Hubble parameter. 
Curves like those in SmurF 4.1 can, for example, always lie above or below -1, with an additional offset of varying magnitude depending on its $w(0)$ value, leading to a posterior very different from the $\Lambda$CDM case.
Intuitively, choosing random $w(0)$ anchoring points leads to a roughly flat distribution of \kld\ values until reaching a maximal possible deviation from $\Lambda$CDM that depends on our allowed $w(0)$ prior range.

\section{Conclusion}
\label{sec:conclusions}

Searching for new physics beyond the standard \lcdm\ model inherently requires the capability to discriminate between competing models for the dark energy equation of state. 
This work scrutinizes the pitfalls of standard cosmological analysis pipelines in their ability to detect signals of \lcdm\ deviations.

For this task, we introduce a novel smooth random curve generator, \smurves, which uses random sampling and modified Newtonian projectile motion as the means for its generative process. 
This method is highly customizable and facilitates the use of physically motivated constraints into the curve-generating process. 
While applied to a specific cosmological case in this paper, \smurves\ represents a general multi-purpose methodology for constrained curve generation and function perturbation. 
We also provide a user-friendly implementation of the code for the sake of reproducible science.

We employ \smurves\ to generate mock \sn\ observations representing four constraint families, or SmurFs, each one representing increasing degrees of deviation from the \lcdm\ model. 
Making use of 50 random $w(z)$ curves per SmurF, we run a Bayesian cosmological inference pipeline for each curve to subsequently produce 200 joint posteriors of $\Omega_{\mathrm{m}}$, $w$, and $M$. 
We then compare these posteriors to those from an analysis of the \pantheon\ sample derived under the assumption of a constant-$w$ model.

We show that \sn\ cosmology observables under extensive redshift dependencies of the dark energy equation of state are virtually indistinguishable from those of \lcdm\ models using current state-of-the-art analysis pipelines. 
Notably, $w(z)$ realizations that exhibit a stronger deviation from $w = -1$ can lead to posterior samples of $\Omega_{\mathrm{m}}$, $w$, and $M$ exhibiting a slightly better agreement with \lcdm\ than realizations with lesser levels of deviation.
This result highlights a fundamental and generally unstated caveat underpinning the current methodology used to estimate $w$ from \sn\ observations: 
If \lcdm\ is assumed as the null hypothesis in a test for compatibility with observational \sn\ data, the inability to rule out the standard model could, in a given case, be based on such similarities in posteriors with potentially large underlying deviations due to statistical degeneracies.

In addition, we test the effect of both an increased number of gradient sign changes, leading to more complex curves, and of larger deviations from $w(z) = -1$ with the omission of an anchor point of $w(0) = -1$ for generated curves.
While the complexity of curves has little impact on the compliance with the standard model, we find that this omission of an anchor constraint at $z = 0$ reduces $\Lambda$CDM compliance considerably.
We recommend further research on the topic, specifically in terms of an investigation focused on different curve characteristics to reduce the set of viable candidate hypotheses.
In doing so, further insights into the specific features of redshift-dependent dark energy equations of state can be gained by identifying regions of $w(z)$ parametrizations that favor certain cosmologies.

The upcoming arrival of larger and higher-quality data sets, especially at high redshifts, will certainly improve our capability to distinguish between dark energy models. 
There are, however, intrinsic characteristics of distance-based observables that can render the identification of strong deviations unattainable. 
The application of redshift-dependent analyses, parametric or non-parametric, alongside the constant-$w$ scenario and the careful use of additional cosmological observables, are crucial steps in providing a realistic picture of our current knowledge regarding properties of dark energy. 
Due to these caveats, and given the significant loss in precision when redshift-dependence is taken into account, physics beyond the standard model may be hidden in plain sight.

\section*{Acknowledgements}

We would like to express our gratitude to Eric D. Feigelson, Arya Farahi, and Alberto Krone-Martins for helpful discussions and suggestions, as well as to Brandon Sanderson for the initial inspiration for the curve generator used in this work.
This work was created during the $\rm 5^{th}$ COIN Residence Program \href{https://cosmostatistics-initiative.org/residence-programs/coin-residence-program-5-chania-greece/}{(CRP\#5)} held in Chania, Greece in September 2018, with support from CNRS and IAASARS.
We thank Vassilis Charmandaris for encouraging the accomplishment of this event. 
This project is financially supported by CNRS as part of its MOMENTUM programme over the 2018--2020 period.
RSS acknowledges the support from NASA under the Astrophysics Theory 
Program Grant 14-ATP14-0007. 
The Cosmostatistics Initiative\footnote{\url{https://cosmostatistics-initiative.org}} (COIN) is a non-profit organization whose aim is to nourish the synergy between astrophysics, cosmology, statistics, and machine learning communities.

\bibliographystyle{apsrev4-1}

\bibliography{stress_testing}

\appendix

\section{Constrained curve generation}
\label{app:smurves}

The segmented path calculation of \smurves\ follows, in its broadest terms, the classical Newtonian calculation of a projectile path: 
Given a velocity, an acceleration magnitude as a force acting on the projectile, and a launch angle, a flight path can be easily computed as vertical axis values along a set of measurement points on the horizontal axis. 
At the end of the partial path computation, the function returns the path measurements, the impact angle, and the final velocity of the projectile. 
Depending on the number of sampled change points, and on whether parts of the full path are not yet calculated, a new force acting in the opposite direction of the previous one is sampled, and previously returned values are re-used as inputs to the same function. 
This lets the projectile continue its flight with the same characteristics, but with changed gravitational magnitude and direction, to ensure a smooth curve evolution that easily lends itself to subsequent splining.

\begin{algorithm}
\label{alg:smurves}
\caption{Partial trajectory calculation}
\KwData{$v :=$ velocity\\
\hspace{30pt}$\alpha :=$ step size\\
\hspace{30pt}$\beta :=$ direction\\
\hspace{30pt}$s :=$ partial steps\\
\hspace{30pt}$p_0 :=$ start point\\
\hspace{30pt}$f :=$ vertical force\\
\hspace{30pt}$\theta :=$ launch angle}
\KwResult{Path $p$, impact angle $\theta_{\textup{imp}}$, velocity $v$}
\textit{Set the initial horizontal displacement to zero}\\
$\Delta x \longleftarrow 0$\\
\textit{Calculate the horizontal and vertical velocities}\\
$v_x \longleftarrow v \cos (\theta)$\\
$v_y \longleftarrow v \sin (\theta)$\\
\textit{Initialize start velocity and path measurements}\\
$v_0 \longleftarrow v$\\
$p \longleftarrow p_0$\\
\textit{Loop over the given x-axis measurement points}\\
\For{$i \gets 1$ \textup{\textbf{to}} $\textup{length}(s)$}{
\hspace{10pt}\textit{Horizontal distance, displacement and time}\\
\hspace{10pt}$d \longleftarrow s[i]$\\
\hspace{10pt}$\Delta x \longleftarrow \Delta x + \alpha$\\
\hspace{10pt}$t \longleftarrow \frac{\Delta x}{v_x}$\\
\hspace{10pt}\textit{Calculate vertical velocity and displacement}\\
\hspace{10pt}$v_y \longleftarrow v_0 \sin (\theta) - f t$\\
\hspace{10pt}$\Delta y \longleftarrow -\left( v_0 \sin (\theta) t - \frac{1}{2} f t^2 \right)$\\
\hspace{10pt}\textit{Total velocity and directional displacement}\\
\hspace{10pt}$v \longleftarrow \sqrt{v_x^2 + v_y^2}$\\
\hspace{10pt}$D \longleftarrow \beta \Delta x$\\
\hspace{10pt}\textit{Append the projectile location at that point}\\
\hspace{10pt}$p \longleftarrow \textup{append}(p, (d, p_0 + D))$
}
\textit{Calculate the impact angle for the partial path}\\
$\theta_{\textup{imp}} \longleftarrow \arctan(-\frac{v_y}{v_x})$\\
\Return $p, \theta_{\textup{imp}}, v$\
\end{algorithm}

The corresponding method for curve segment calculations is specified, as pseudocode, in Algorithm~\ref{alg:smurves} to allow for an easier replication and easier understanding both of our approach and the accompanying open-source code implementation for smooth random curve generation.

\section{Parameter estimation with \emcee}
\label{app:emcee}

For our parameter estimation, we employ \emcee, a popular pure-Python implementation of the affine-invariant MCMC ensemble sampler. 
This approach extends the classic Metropolis-Hastings algorithm with a parallel ``stretch move''. 

A number $K$ of walkers explore the parameter space, with their respective steps drawn from a proposal distribution that depends on other walkers' positions. 
A walker at position $Y$ is drawn by chance to propose a new position $X'$ for the walker that is to be updated and currently at position $X$, meaning that
\begin{eqnarray}
X \rightarrow X' = Y + Z[X - Y].
\end{eqnarray}
Here, $Z$ acts as a random variable with $S \coloneqq \left[ 0.5, 2 \right]$ and $Z \sim g(z) \propto \mathbf{1}_S(z) \cdot {\sqrt{z}}^{-1}$, with the indicator function $\mathbf{1}_S(z)$ taking a value of one for all $z \in S$ and a value of zero for all $z \notin S$. 
Alternatively, this can be written as
\begin{eqnarray}
g(z) \propto
\begin{cases}
\frac{1}{\sqrt{z}} & \textup{if } z \in \left[ \frac{1}{2}, 2 \right] \\
0 & \textup{otherwise}
\end{cases}.
\end{eqnarray}
The ``parallel stretch'' mentioned above splits the $K$ walkers into two equal-sized subsets and updates all walkers of one subset using the other, followed by the corresponding opposite procedure, which allows for the parallelization of this computationally expensive update step.

An affine-invariant MCMC algorithm satisfies $X_a(t) = AX_b(t) + b$ for different starting points $X_a$ and $X_b$, and two probability densities $\pi$ and $\pi_{A,b}$, for any affine transformation $Ax + b$.
The independence of the aspect ratio in highly anisotropic distributions offers a speed advantage in highly skewed distributions.

\section{\cosmosis\ noise addition and bug fix}
\label{app:cosmosis}

\cosmosis\ generates simulations of peak B-band magnitudes as $m_\mathrm{B} (z_i)$ double arrays based on binned \pantheon\ \sn\ data. 
From the \pantheon\ noise covariance $C \equiv \braket{nn^T}$, we can generate this simulation using its (unique) Cholesky decomposition $C = LL^T$ and a random vector $r$, where each element is a random normal value with $r_i \sim N(0,1)$. 
We can then form $n = L \cdot r$ as our noise simulation, as the noise covariance is then
\begin{eqnarray}
\braket{nn^T} = \braket{Lrr^TL^T} = \braket{LL^T} = C.
\end{eqnarray}
As a consequence, the total simulated values $m_{\mathrm{B}_{\textup{sim}}}$ obtained through \cosmosis\ are
\begin{eqnarray}
\label{eq:covariance}
m_{\mathrm{B}_{\textup{sim}}} = m_{\mathrm{B}_{\textup{truth}}}(z_i) + L \cdot r,
\end{eqnarray}
for true values $m_{\mathrm{B}_{\textup{truth}}}$.
Initial experiments to compare the original Pantheon data with \sn\ data generated using flat $w(z)$ curves as a null test uncovered a bug in \cosmosis. 
After this was reported and subsequently fixed, the flat-curve simulations of \sn\ peak B-band magnitudes returned to expected values of $m_\mathrm{B}$.

\section{Interpretation of posterior samples}
\label{app:math}

Given the way in which posteriors from $w(z)$ curve realizations from the same constraint family are used in this paper, one might ask why posterior samples obtained from instances of the same SmurF are not simply combined to arrive at a posterior for the constraint family. Considering error contours as being comprised of samples from $p(\theta | D, \Psi)$, as introduced in Section~\ref{sec:metric}, neglects the role of the initial conditions $C_{0}$ that have been implicitly marginalized out as
\begin{eqnarray}
\label{eq:appmarginalize}
    p(\theta | D, \Psi) = \int p(C_{0}, \theta | D, \Psi) d C_{0} .
\end{eqnarray}
Since we generally cannot constrain the initial conditions as such, an obvious question to ask is why they matter.

When combining constraints on cosmological parameters from different probes $D$ and $D'$, we are really asking for $p(\theta | D, D', \Psi)$ when we have $p(\theta | D, \Psi)$ and $p(\theta | D', \Psi)$.
To make use of the independence of the datasets, we would expand this in terms of Bayes' Rule as 
\begin{eqnarray}
\label{eq:appbayes}
\hspace{-20pt} p(\theta | D, D', \Psi) &=& \int p(C_{0}, \theta | D, D', \Psi) d C_{0}\\
    &=& \int p(D, D' | C_{0}, \theta, \Psi) \frac{p(C_{0}, \theta | \Psi)}{p(D, D' | \Psi)} d C_{0} .
\end{eqnarray}
If $D$ and $D'$ are our standard independent probes, every term in Equation~\ref{eq:appbayes} is well-defined. This means that the integral is separable and we can recover the intuitive way to combine the posteriors.

The situation investigated in this paper, however, is different.
In our case, $D$ and $D'$ correspond to different SmurF instances $j$ and $j'$.
These two datasets are inherently contradictory; they could never be observed in the same instantiation of the universe, even under the same physical model and values of the cosmological parameters $\theta$.
In other words, $p(D, D' | C_{0}) = 0$ for any pair of mock-\pantheon\ data we consider.
What distinguishes one SmurF from another is rolled into the initial conditions $C_{0}$, leading to well-defined $p(\theta | D, \Psi)$ and $p(\theta | D', \Psi)$, but to an internally inconsistent $p(\theta | D, D', \Psi)$.
Thus, it would be inappropriate to combine samples  of the cosmological parameters obtained through a Markov chain Monte Carlo (MCMC) method from any collection of SmurF instances with different $w(z)$ curves, divided by constraint family or not.

\end{document}